\documentclass[12pt]{iopart}
\usepackage{graphicx}

\newcommand*\VF[1]{\mathbf{#1}}
\newcommand*\dif{\mathop{}\!\mathrm{d}}

\begin{document}

\title{Levitation, oscillations, and wave propagation in a stratified fluid}

\author{Marina Carpineti$^1$, Fabrizio Croccolo$^2$, and Alberto Vailati$^1$}                     
%

\address{$^1$Dipartimento di Fisica Aldo Pontremoli, Universit\`{a} degli Studi di Milano, I-20133 Milano, Italy }

\address{$^2$Universite de Pau et des Pays de l'Adour, E2S UPPA, CNRS, TOTAL, LFCR UMR5150, Anglet, France.}
\ead{marina.carpineti@unimi.it}
\vspace{10pt}

\date{\today}

\begin{abstract}
	
	We present an engaging levitation experiment  that students can perform at home or in a simple laboratory using everyday objects. A cork, modified to be slightly denser than water, is placed in a jug containing tap water and coarse kitchen salt delivered at the bottom without stirring. The salt gradually diffuses and determines a stable density stratification of water, the bottom layers being denser than the top ones. During the dissolution of salt, the cork slowly rises at an increasing height, where at any instant its density is balanced by that of the surrounding water. If the cork is gently pushed off its temporary equilibrium position, it experiences a restoring force and starts to oscillate.
Students can perform many different measurements of the phenomena involved and tackle non-trivial physical issues related to the behaviour of a macroscopic body immersed in a stratified fluid. Despite its simplicity, this experiment allows to introduce various theoretical concepts of relevance for the physics of the atmosphere and stars and offers students the opportunity of getting acquainted with a simple system that can serve as a model to understand complex phenomena such as oscillations at the Brunt-V\"{a}is\"{a}l\"{a} frequency and the propagation of internal gravity waves in a stratified medium.

\end{abstract}

\maketitle 

\section{Introduction}

The levitation of macroscopic objects has fascinated humankind for a very long time. Its history is intertwined with the research performed by several Nobel laureates, because beyond its mystical aspects levitation can be profitably used both as a tool in fundamental research and in technological applications \cite{Brandt89,Callens2011}. Beyond its historical role in science, the investigation of levitation is still full of puzzling results, such as the apparent reversal of the gravity acceleration recently reported for a solid body at the surface of a levitating liquid \cite{Sorokin2020,Apffel2020}.

The fascination determined by levitation comes from the fact that we rarely experience it, as objects close to the surface of the Earth do not float freely into the air, due to the strong gravitational attraction and the small buoyancy force exerted by air onto solid bodies. 
When the gravitational force is not balanced, the object is in free fall, and in the frame of reference of the falling object, the acceleration of gravity is not felt. This determines weightlessness levitation conditions that can be employed profitably to perform experiments in the absence of gravity on platforms such as drop towers, parabolic flights, and artificial satellites like the International Space Station \cite{Braibanti2019}.
On Earth, the gravitational force acting on a body can be also balanced by the buoyancy force acting onto it when the body is immersed in a fluid medium. When the density of the body equals that of the surrounding fluid, the weight of the body is exactly balanced by the buoyancy force and the body is in equilibrium inside the fluid neither rising nor sinking, in a condition similar to levitation.  However, the phenomenon of levitation does not involve simply the balance of forces, as occurs in buoyancy phenomena, but the mechanical stability of the levitating body \cite{Jones1997}. The particularity of a levitation phenomenon  is that restoring forces drive back the  body when it is moved away from its equilibrium position.

A levitation condition can occur in a stable stratified medium, where the density decreases as a function of height. A body can therefore  be in equilibrium only at a certain height where its density is exactly matched and a displacement in a direction parallel to the density gradient determines a restoring force that brings back the object to the layer of fluid with the same density. This force gives rise to damped harmonic oscillations that occur at a frequency, known as Brunt-V\"{a}is\"{a}l\"{a} frequency \cite{nappo},  which is determined by the local density gradient present in the stratified fluid. The oscillations gradually damp both through viscous dissipation and through the emission of transverse internal gravity waves propagating in the plane perpendicular to the direction of oscillation.  Examples of natural systems where a density stratification occurs include giant planets and stars \cite{LeBars2020}, where the mass is large enough to determine the confinement of a fluid phase. A typical example is represented by the atmosphere of a planet, where the stratification of a gaseous phase is due to the variation of pressure with altitude, which determines a gradual decrease of density.

In this work, we discuss a simple experiment that can be realized by students using kitchenware to investigate levitation,  oscillations at the Brunt-V\"{a}is\"{a}l\"{a} frequency, and internal gravity waves in a stratified medium. The experiment makes use of a modified cork that sinks into water but suddenly starts levitating when a large enough saline gradient is generated by pouring coarse salt inside water. Its simplicity makes this demonstration suitable for students starting from elementary school up to undergraduate school, due to the opportunity of achieving different levels of understanding and modelling of the physical processes involved. 

For undergraduate students, in particular, the experiment can be used to convey the hydrostatic principles behind the balance of the forces involved. Moreover, an external perturbation of the cork allows the investigation of the physics of oscillations in a stratified medium and the radiative transfer of the energy of oscillations determined by internal gravity waves.

The paper is organized as follows. The first paragraph reviews the experimental contexts where levitation is observed in nature, in order to give a general framework where the experiment may be inserted. In the following, we present the experiment, discuss its relevance in an educational context, and describe the experimental materials and methods. We then provide the basic theoretical concepts needed to describe the hydrostatic equilibrium of a fluid and the conditions that can give rise to either buoyancy or levitation of a body embedded in it. We discuss the case of the oscillations of a body in equilibrium in a stratified fluid and the internal gravity waves originated by the dissipation of energy during the oscillations. Finally, we describe, analyze and discuss the experimental results.

\section{Levitation in science}

Although it hardly occurs spontaneously in nature, levitation has been profitably used in science to perform several challenging experiments. A notable example of its usage in fundamental research is represented by the celebrated experiment by Robert Millikan for the measurement of the charge of the electron, where a negatively charged droplet of oil was levitated by using an electrostatic field \cite{Millikan1913}. Millikan was awarded the Nobel prize in Physics in 1923 for this experimental demonstration of the discrete nature of the electric charge \cite{Millikan1924NL}.  
Levitation can also be achieved by using a magnetic field, as it happens for example by placing a superconducting disk above a strong magnet \cite{Landau60, Saslow90, Brandt89}.   Magnetic levitation has important technological applications for the development of frictionless railways \cite{Powell71, Coffey94}. A strong magnetic field can also be used for challenging applications like the levitation of a living being, as it has been done in a popular experiment by Andre Geim and Michael Berry. By using a strong magnetic field, they were able to levitate a living frog, and this result allowed them to win the Ig-Nobel prize in Physics in the year 2000 \cite{Berry1997}.  A few years later, in 2010, Geim shared the Nobel prize in Physics with Konstantin Novoselov for the discovery of graphene \cite{Geim2010NL}.
Beyond the use of external fields, levitation can be also achieved by using radiation \cite{Brandt89}. The phenomenon of acoustic levitation was studied by Rayleigh. Currently, acoustic levitation can be even achieved by amateurs at home or in the classroom, by building affordable instrumentation based on off-the-shelf components such as Tiny-Lev \cite{Marzo2017}, a device that allows levitating  small objects with a maximum size of 4 mm and density below 2 g/cm$^3$. The levitation of particles using light can be achieved by using optical tweezers, developed by Arthur Ashkin \cite{Ashkin1970}, an invention that allowed him to win the Nobel Prize in Physics in 2018 \cite{Ashkin2018NL}. Optical tweezers allow the non-invasive manipulation of nanoparticles by taking advantage of a focused Laser beam that exerts a non-isotropic radiation pressure, resulting in the trapping of a particle in a potential well. The introduction of optical tweezers determined a revolution in the fields of Biophysics and Soft Matter, where they allow the investigation of the behaviour of colloidal particles, molecular motors, and cells, to mention just a few representative applications \cite{Moffit2008, McGloin2010}. 

Neutral buoyancy represents a condition similar to levitation,  very common in the case of animals populating the waters of the Earth. Meaningful strategies adopted by marine animals to achieve neutral buoyancy are represented by the swim bladder of teleost fish, by the squalene oil contained in large quantities into the liver of sharks, and by the large hollow chambers built by the nautilus into its shell \cite{Vogel2013}. In the case of marine animals, neutral buoyancy is achieved in a medium of nearly constant density, and this feature determines an indifferent equilibrium condition, where the weight of the animal can be balanced by buoyancy irrespectively of the position of the animal. The lack of a unique equilibrium position and of a restoring force indicates that neutral buoyancy cannot properly be considered a levitation phenomenon.

At variance, the equilibrium condition realized in the presence of a stable density stratification can be well described as levitation. In particular, when the density gradient evolves in time, it is possible to observe that an object, initially at rest, suddenly starts to rise under the action of invisible forces. At any instant, its equilibrium is at a well-defined height where it is subjected to restoring forces that make it oscillate in case it is slightly displaced. 
Several macroscopic systems exhibit these conditions and share the common feature that the mechanical energy associated with the local oscillation can be transferred at a distance by internal gravity waves generated by the oscillation itself. Although internal gravity waves in a fluid cannot be visualized, the effect of the energy transferred by them at a distance can be appreciated from the effect that it has on macroscopic bodies embedded in the fluid.
 An important example is represented by the Earth's atmosphere, where mountain reliefs and thunderstorms can interact with gravity waves to give rise to complex turbulent phenomena \cite{nappo}. Another notable example is represented by oceans, where the density stratification is both determined by changes in salinity and temperature.
Internal gravity waves have been reported also at the mesoscale in laboratory experiments on non-equilibrium fluids, where the displacement of parcels of fluids is determined by the thermal energy $k_BT$ \cite{Croccolo2019}. 

\section{The experiment} 
The idea of this experiment came during the lockdown period due to the COVID-19 pandemic when forced to stay far from our laboratories we were stimulated to invent simple and catching experiments, easy to be performed at home but able at the same time to teach physics.
The experiment is ideally presented to students through an inquiry based path, where the phenomenology is first demonstrated to the students by a facilitator. Students are then instructed about the details of the experiment so that they can perform it autonomously. Eventually, the students are lectured about the physical principles behind the experiment and about the methods for the analysis of results.

\subsection{Inquiry-driven presentation of the experiment}

\begin{figure}[h!]
	\centering
	\includegraphics[width=16.5 cm]{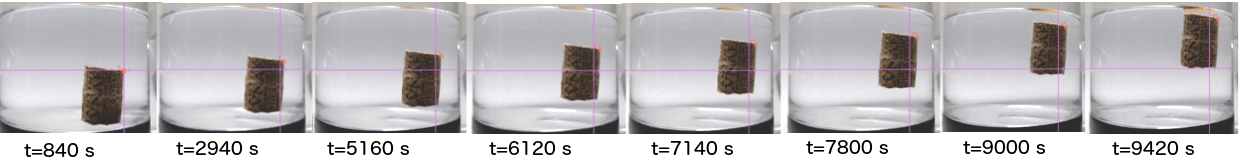}
	\caption{Shots of the cork during its rise. The marked point and the reference axis used for tracking are shown. Coordinates are fixed so that the origin coincides with the position of the tracked point at the starting time. For each figure, the time of acquisition from the beginning of the measurement is shown. }
	\label{fig:cork_rise}
\end{figure}

Levitation is a surprising, intriguing, and counter-intuitive theme that offers the chance to discuss many physical phenomena. It is extremely difficult to observe it in practice and all the experiments involving it require a great technological effort. 
In this work, we describe a very simple levitation experiment, where a weighed-down cork,  initially located at the bottom of a glass jug containing water, suddenly starts levitating inside the fluid and gradually rises (Fig. \ref{fig:cork_rise}). 

\begin{figure}[h!]
	\centering
	\includegraphics[width=16.5 cm]{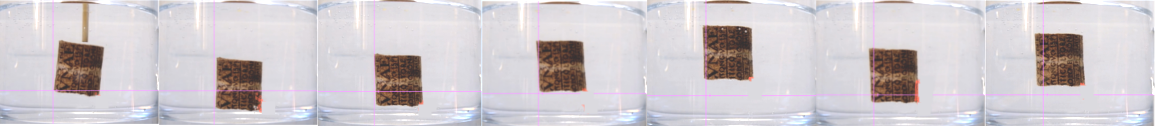}
	\caption{Shots of the cork during oscillation induced by a slight initial push. A video is included as supplemental material (video1).}
	\label{oscillation}
\end{figure}

  Apparently, no source of energy is involved in the process, and this leaves the students with a first scientific puzzle: \textit{i) where does the mechanical work needed to raise the cork come from?}
Another striking aspect of this experiment is that when the levitating cork is slightly displaced vertically from its position, it undergoes damped harmonic oscillations (Fig. \ref{oscillation}). This aspect leaves the students with a second puzzling question: \textit{ii) what is the physical origin of the restoring force that drives the cork towards its equilibrium  position during oscillations?}

\begin{figure}[h!]
	\centering
	\includegraphics[width=6 cm]{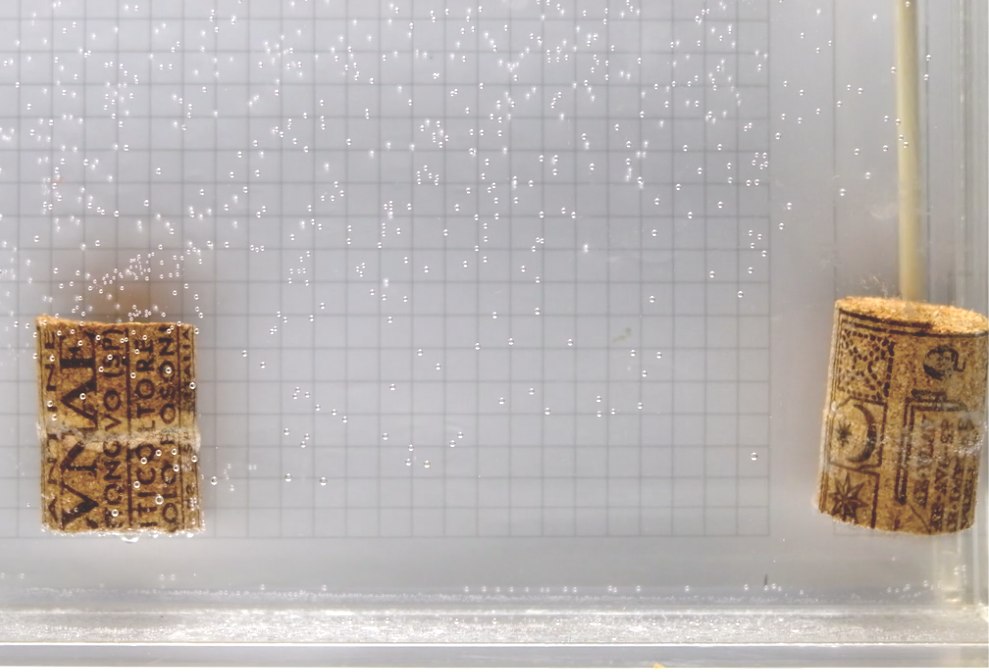}
	\caption{Two corks in a tank. After some oscillations of the right cork, also the left cork starts to move although with an oscillation amplitude much lower than the first cork. A video is included as supplemental material (video2).}
	\label{Two_corks}
\end{figure}

When two levitating corks are hosted inside the same container, pushing one of the corks vertically determines, after a latency time, the onset of oscillations in the second cork as well  (Fig. \ref{Two_corks}). This feature raises a third puzzling question for the students: \textit{iii) what is the origin of the interaction between the two corks?}

\subsection{Materials and methods}

\begin{figure}[h!]
	
	\centering
	\includegraphics[width=6cm]{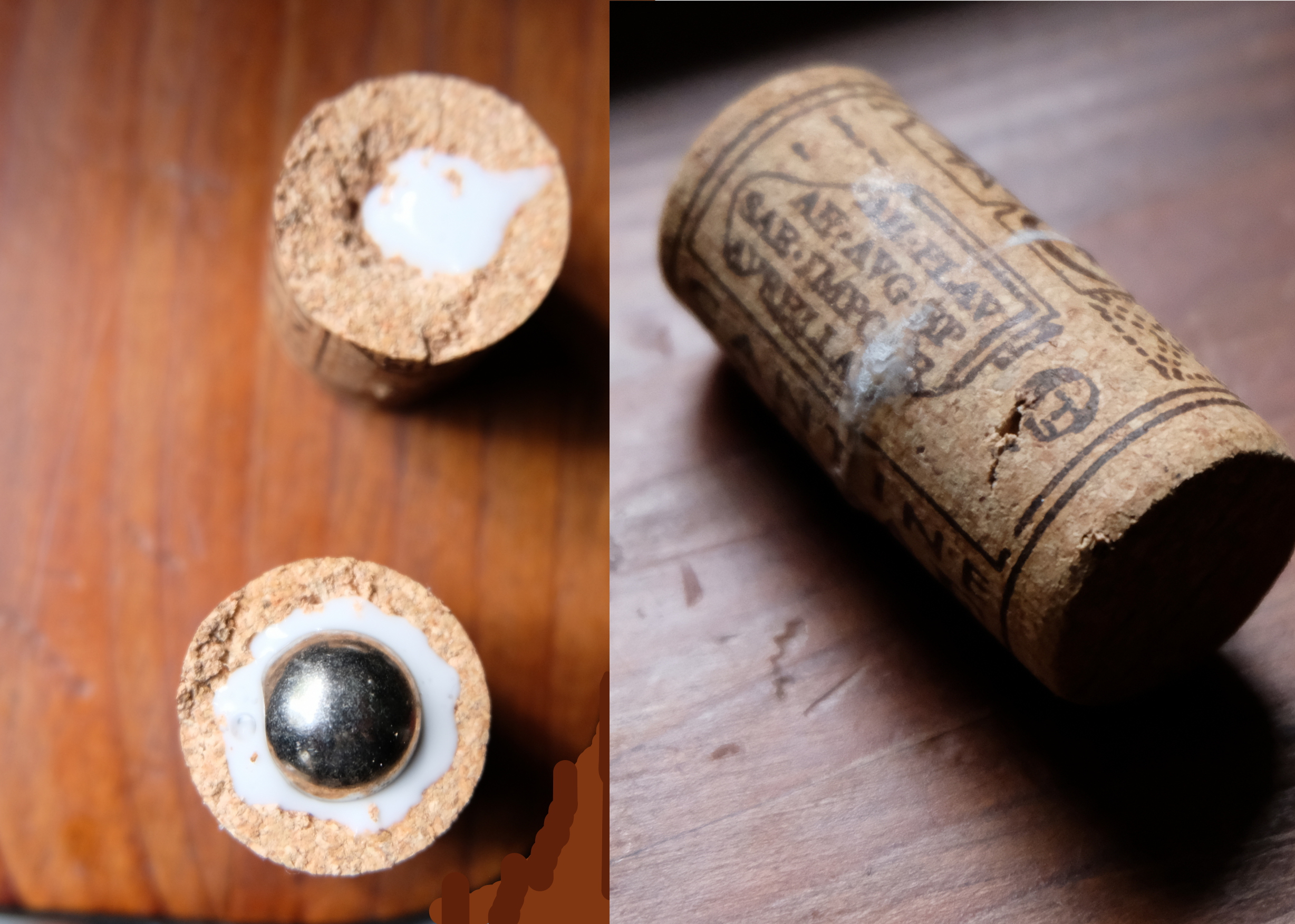}
	\caption{The picture on the left shows the two halves of the cork, with the hemispherical dimples carved into them, both filled with vinyl glue and one of them hosting the metal sphere. The picture on the right shows the assembled cork. In this configuration, the cork density was still smaller than that of water and it was necessary to cut thin slices of the cork to make it slightly denser than water.}
	\label{fig:cut_cork}
	
\end{figure}

For the experiment we had in mind, we needed an easy-to-find object, with a density slightly larger than water. The choice was to weigh-down the cork of a wine bottle. The cork was cut in two halves digging a hole in each of them to host a weight, for which we chose another item frequently available at home: a steel marble with a diameter of $ 10 \ mm$ borrowed from a toy. We soaked the sphere in vinyl glue, pasted back the two halves of the cork (see Fig. \ref{fig:cut_cork}), and used ethyl cyanoacrylate glue along the junction. Finally, we cut thin slices of the cork, until it had a density slightly larger than water and was able to sink into it.

The experiment is prepared by pouring  $500 \ ml$ of tap water into a cylindrical glass jug, and adding a defined amount of course kitchen salt without applying any stirring, so that the salt grains rapidly sediment to the bottom of the jug.  We decided to use coarse rather than fine salt to avoid its partial dissolution while settling at the bottom of the jug, and therefore guarantee to start from better-defined initial conditions. The experiment is started by rapidly putting the cork at the bottom before the salt has much time to dissolve. The cork, initially sitting at the bottom of the jug, is expected to start to levitate when the surrounding density becomes higher than its own and to move towards a height where an iso-density condition is reached. The process occurs because the salt undergoes a diffusion process that determines a density stratification in the water: the salt-rich layers of fluid at the bottom of the jug are initially denser than the salt-poor layers at the top. Diffusion determines the spread of the salt ions across water, until eventually they become uniformly distributed inside the available volume.

Once the density of the cork is chosen, measurements can be performed with different amounts of salt that determine the behaviour of the cork. We span densities varying from 10 to 46 g/l.

In our experiment, for the smaller amounts used (typically 10-15 g/l), the cork rises after some time but eventually goes back to the bottom of the jug, because the equilibrium density (i.e. for infinite time) of the aqueous solution is smaller than that of the cork. For larger amounts of salt, the cork rises at the beginning and eventually reaches the top of the liquid, finally floating at its surface. In the latter case, the equilibrium density of the liquid is larger than that of the cork. Refining the quantity of salt dissolved in the tap water around the values obtained for the two different conditions, one can get a somewhat precise measurement of the density of the cork. We could estimate its density to be approximately $ 1,030 \ g/cm^3$, starting from an estimation of the critical amount of salt of 40 g dissolved in 1 kg of tap water, although the cork density can slightly change from one experiment to another due to its moisture content.

The cork ascent lasts usually many hours and we recorded it using a Fujifilm XT20 camera programmed to shoot pictures at regular time intervals, typically every five minutes. The framing is chosen so as to capture the entire height of the liquid.
In order to study the oscillation process, we apply a punctual acceleration to the cork and record its oscillations at 24 frames/second by using the same  camera. 
Should this experiment be proposed to students, even a readily available smartphone could be used to make many of the measurements and this was indeed the procedure followed at the beginning. However, when available, an automated system is definitely preferable.

\begin{figure}[h!]
	
	\centering
	\includegraphics[width=6cm]{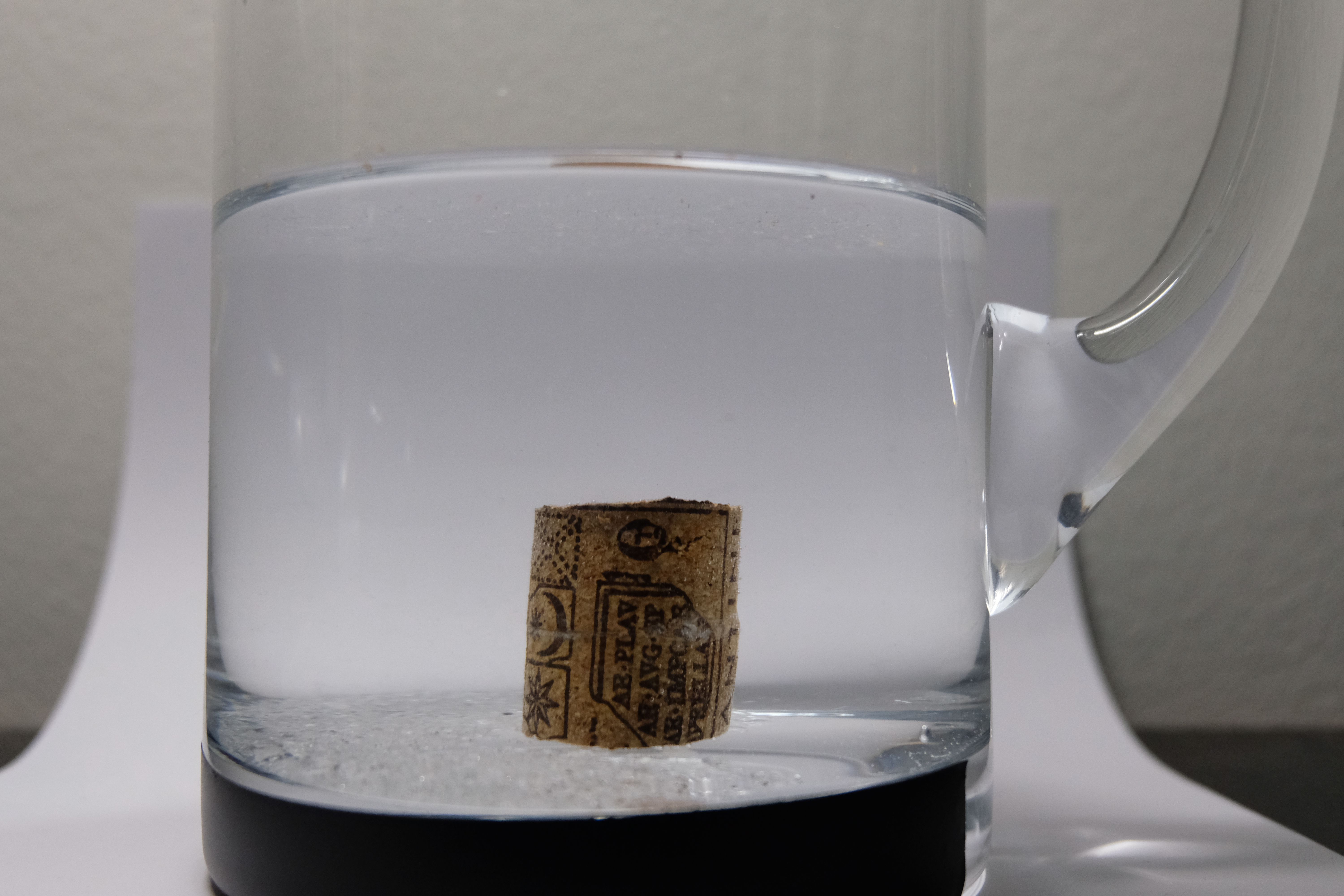}
	\caption{Picture of the starting time of a typical measurement. The cork lies at the bottom of the jug where the coarse salt has settled.}
	\label{fig:exp}
\end{figure}

Both cork's rise and oscillations were characterized using the Open Source software "Tracker" \cite{linkTracker}, a Java-based application that allows to track the different positions of an object in a video. To analyse the cork's ascent, the sequence of collected images has been first converted in an AVI video, using the  Fiji Open Source software \cite{linkFiji}, and then imported into Tracker, setting the correct frame rate. 
Spatial calibration is done for each video by assigning to the water height the real value previously measured with a ruler ($6.8 \ cm$). To track the different positions of the cork, a mark is then placed at a recognizable point of it, such as a corner or a small detail, and the origin of the reference system is fixed on it. Due to a large number of frames to analyse, the choice of a well-identifiable and non-ambiguous point is extremely important to allow the use of the auto-tracking procedure, much faster than the manual one.  At the end of the process, the different positions of the point as a function of time are obtained.

\section{Theory review} 

In this section, we will provide the basic physical concepts needed for the understanding of the hydrostatic equilibrium of a fluid, which can give rise to the levitation of a body embedded into it and, consequently, to its oscillations when it is gently pushed off its equilibrium position. The theory of internal gravity waves will be also briefly outlined.

\subsection{Hydrostatic equilibrium, buoyancy, and levitation} 

Let us consider a very general configuration suitable to describe fluids such as the atmosphere and oceans: a layer of fluid of density $\rho$, distributed in a spherical shell of thickness $L$, located ar a distance $R$ from the centre of mass of a distribution of total mass M. 

A parcel  of fluid of volume V is under the action of the gravity force $F=\rho V g$ determined by the Law of Universal Gravitation $F=G\frac{mM}{R^2}$, where the acceleration of gravity is given by $g(R)=G\frac{M}{R^2}$. At a height $z$,  the weight of the column of fluid above it determines a hydrostatic pressure $p$ inside the fluid, governed by the equation:

\begin{equation}
\nabla p= \rho \, \VF{g} 
\label{eq:hydpress}
\end{equation}

In the case of an incompressible fluid like a liquid, the density is constant and the hydrostatic pressure is governed by Stevino's equation 

\begin{equation}
p(z)=p_0-\rho g z
\label{eq:stevino}
\end{equation}

which is a particular solution of Eqn. \ref{eq:hydpress}, where $p_0$ is the pressure at the surface of the fluid $z=0$. 

A consequence of the hydrostatic pressure of Eqn. \ref{eq:hydpress} is that the surface $S$ of a body of volume $V_0$ completely immersed in the fluid is subjected to forces acting perpendicularly to its surface. As the horizontal components are balanced,  the body is under the action of a buoyancy force $\VF{F}_A$ directed vertically, which can be obtained by integrating the hydrostatic pressure provided by Eqn. \ref{eq:stevino} across the surface $S$ of the body:

\begin{equation}
{F}_A =-\left| \oint_{S} p(z) \dif \VF{S}\right|  =\int_{V_0} \frac{dp}{dz} \dif V =  \rho \, {g} \, V_0
\label{eq:gauss}
\end{equation}

where $\dif\VF{S}$ is an element of the surface with outgoing normal, and we have applied the divergence theorem to convert a surface integral to a volume integral. 

The total force acting on the body can be obtained by adding to the buoyancy force the weight of the body $\rho_0 \, V_0 \, {g}$ with the proper sign:

\begin{equation}
{F}_T =\left(\rho-\rho_0\right)  \, {g} \, V_0
\label{eq:buoyancy_tot}
\end{equation}

where $\rho_0$ is the density of the body.
An immediate consequence of Eqn. \ref{eq:buoyancy_tot} is that when the density of the body is matched with that of the fluid,  $\VF{F}_T =0$. Under these conditions, the body is in an indifferent equilibrium condition called neutral buoyancy, i.e the equilibrium condition is achieved at each point of the fluid. Such condition is achieved by several aquatic organisms, which have devised a series of strategies to compensate for the variation of pressure as a function of depth \cite{Vogel2013}. A major limitation of neutral buoyancy is that it requires a very close matching between the density of the fluid and that of the body. Even a small difference between these quantities would give rise to a condition where the body is under the action of a constant acceleration, which brings it progressively farther from its initial position. 

A stable equilibrium condition can be achieved by immersing the body in a stratified compressible fluid, like for example a gas stratified by the gravity force. Under this condition, the body, after a transient, migrates to the layer of fluid matching its density. This condition is difficult to obtain in practice because the average density of most solid bodies is much larger than that of a gas. Matching the density of the solid would require the use of a liquid, which is intrinsically not compressible and, therefore, cannot be stratified by the gravity force.

Notwithstanding that, a gravitationally stable density stratification can be obtained in a liquid under non equilibrium conditions. In fact, if the liquid is under the action of a vertical temperature gradient $\nabla T$  its thermal expansion determines a density gradient $\nabla \rho = \frac{\partial \rho}{\partial T} \nabla T$. Moreover, if the liquid is a mixture of two components, a variation of the concentration $c$ with height determines a density gradient $\nabla \rho = \frac{\partial \rho}{ \partial c} \nabla c$. In general, combining these two terms, the overall vertical density gradient is given by:

\begin{equation}
\nabla \rho = \rho \left( \alpha  \nabla T + \beta \nabla c\right) 
\label{eq:bouss}
\end{equation}

where $\alpha$ is the thermal expansion coefficient, and $\beta$ is the solutal expansion coefficient. 

Under the condition that the overall density gradient points downwards, the density profile is gravitationally stable, and a body of density $\rho_0$ such as  $\rho_{min}<\rho_0<\rho_{max}$ can levitate in a layer of a stratified fluid.

\subsection{Oscillations in a stratified fluid}

Let us now consider the case where a body of density $\rho_0$  and volume $V_0$ is immersed in a layer, with the same density, of a gravitationally stable  stratified fluid. If the volume $V_0$ undergoes a small displacement $z$ in the vertical direction, the density gradient can be assumed to be locally constant so that the density changes linearly with height:

\begin{equation}
\rho(z) = \rho_0 + \frac{\partial \rho}{ \partial z} \, z
\label{eq:densprof}
\end{equation}

Combining Eqns. \ref{eq:hydpress}, \ref{eq:gauss}, and \ref{eq:densprof}, one can calculate the buoyancy force acting on volume $V_0$:

\begin{equation}
{F}_B  = -\oint_{S} p(z)   \dif \VF{S} =\int_{V_0} \frac{dp}{dz} \dif V =  \rho_0 \, {g} \, V_0 + \frac{\partial \rho}{ \partial z} \, {g} \, V_0 \, z
\label{eq:harmonic}
\end{equation}

and recalling Eqn. \ref{eq:buoyancy_tot} the total force acting on $V_0$ is:

\begin{equation}
{F}_T = \frac{\partial \rho}{ \partial z} \, {g} \, V_0 \, z
\label{eq:buoyancy_tot_strat}
\end{equation}

By applying the second principle of dynamics $\rho_0\, V_0 \, \ddot{z}=F_T$, one obtains the equation of motion of a harmonic oscillator (notice that $\frac{\partial \rho}{ \partial z}$ is negative) :

\begin{equation}
\ddot{z} = \frac{g}{\rho_0}\frac{\partial \rho}{ \partial z}  \, z
\label{eq:eq_mot_harm}
\end{equation}

where oscillations occur at the Brunt-V\"{a}is\"{a}l\"{a} frequency \cite{nappo}:

\begin{equation}
N= \sqrt{-\frac{g}{\rho_0}\frac{\partial \rho}{ \partial z}  }
\label{eq:brunt}
\end{equation}

Oscillations at the Brunt-V\"{a}is\"{a}l\"{a} frequency also occur when a parcel of fluid is displaced vertically with respect to the surrounding fluid. This process occurs in  natural systems such as the atmosphere and the oceans and in astrophysical systems such as stars. Interestingly, the same physical mechanism drives oscillations at mesoscopic length scales during a thermal diffusion process occurring in a mixture of fluids \cite{Croccolo2019}. In that case, the density stratification stems from temperature and concentration gradient, thus resulting in an additional contribution related to the temperature gradient. Anyway, the equation can be written in the same form of Eq.\ref{eq:brunt}  of the present paper if one considers the dependence from the density gradient.

 \subsection{Internal gravity waves}
 
When the volume $V_0$ oscillates, its mechanical energy is gradually transferred to the surrounding fluid. Part of the energy undergoes viscous dissipation, while another part gets transferred at a distance by the layered fluid in the form of an internal gravity wave. Internal gravity waves are qualitatively similar to the surface waves in a liquid. However, their density gradient is, of course, much smaller than that of surface waves. In the following we will describe the emission of transverse waves by a body that undergoes vertical harmonic oscillations at the Brunt-V\"{a}is\"{a}l\"{a}  frequency in a stratified fluid . For a more rigorous and general treatment of internal gravity waves see Tritton \cite{tritton} and Nappo \cite{nappo}.

We consider an incompressible inviscid fluid in the presence of a stable density stratification described by Eqn. \ref{eq:densprof}. The Eulerian hydrodynamic equations for a fluid of velocity $\VF{u}$, pressure $p'$ and density $\rho'$ are:

\begin{eqnarray}
\label{eq:continuity}
\nabla \cdot \VF{u}=0 \\
\label{eq:momentum}
\rho'\, \frac{\partial \VF{u}}{\partial t}+ \rho' \, \VF{u} \cdot \nabla  \VF{u}= - \nabla p' + \rho' \, \VF{g} \\
\label{eq:density}
\frac{\partial \rho'}{\partial t}+ \VF{u} \cdot \nabla \rho'= 0
\end{eqnarray}

Equation \ref{eq:continuity} is the continuity equation and expresses the conservation of mass, while Eqn. \ref{eq:momentum} is the second law of dynamics and Eqn. \ref{eq:density} expresses the fact that the fluid is not compressible.

We assume that the variables for the fluid in motion with velocity $\VF{u}$ can be written as a superposition of an hydrostatic contribution and a small perturbation:

\begin{eqnarray} 
\label{eq:fluctuations}
p'(x,y,z,t)= p(z)+\delta p(x,y,z,t)\\
\label{eq:fluctuations2}
\rho'(x,y,z,t)= \rho(z)+\delta \rho(x,y,z,t)
\end{eqnarray}

where $p$ and $\rho$ obey the hydrostatic Eqns. $\nabla p= \rho g$ and $\partial \rho/\partial t=0$, respectively. By inserting Eqns. \ref{eq:fluctuations} and \ref{eq:fluctuations2} into Eqns \ref{eq:continuity}-\ref{eq:density}, we obtain:

\begin{eqnarray} \label{eq:continuity1}
\nabla \cdot \VF{u}=0 \\
\label{eq:momentum1}
\rho\, \frac{\partial \VF{u}}{\partial t}+ \delta \rho\, \frac{\partial \VF{u}}{\partial t}+ \rho \, \VF{u} \cdot \nabla  \VF{u}+ \delta \rho \, \VF{u} \cdot \nabla  \VF{u}= - \nabla \delta p + \delta\rho \, \VF{g} \\
\label{eq:density1}
\frac{\partial \delta \rho}{\partial t}+ \VF{u} \cdot \nabla \rho +\VF{u} \cdot \nabla \delta \rho= 0
\end{eqnarray}

We now assume that the perturbations of density and pressure are small with respect to the hydrostatic quantities, $|\delta p/p|\ll1$ and $|\delta \rho/\rho|\ll1$, and linearize the equations in $\VF{u}$, $\delta p$, and $\delta \rho$:

\begin{eqnarray} \label{eq:continuity_lin}
\nabla \cdot \VF{u}=0 \\
\label{eq:momentum_lin}
\rho\, \frac{\partial \VF{u}}{\partial t}= - \nabla \delta p + \delta\rho \, \VF{g} \\
\label{eq:density_lin}
\frac{\partial \delta \rho}{\partial t}+ \VF{u} \cdot \nabla \rho = 0
\end{eqnarray}

We look for solutions of the equations in the form of propagating harmonic waves of wave vector $ \VF{k}$ and frequency $\omega$:

\begin{eqnarray} \label{eq:wave_velocity}
\VF{u}=\VF{U} \,exp\left[i\left(  \VF{k}\cdot  \VF{x}-\omega t\right)  \right] \\
\label{eq:wave_density}
\delta \rho=R \,exp\left[i\left(  \VF{k}\cdot  \VF{x}-\omega t\right)  \right]  \\
\label{eq:wave_pressure}
\delta p=P\,exp\left[i\left(  \VF{k}\cdot  \VF{x}-\omega t\right)  \right] 
\end{eqnarray}

For simplicity, we restrict ourselves to a two-dimensional system, of coordinates $x$ (horizontal), and $z$ (vertical). We assume that the parcel of fluid is located in the origin of the coordinate system at time $t=0$, and undergoes harmonic motion in the vertical direction. This implies that the amplitude $\VF{U}$ of the oscillation in Eqn. \ref{eq:wave_velocity} must be directed in the vertical direction, that is $\VF{U}=(0 \ ; U)$.

By imposing that Eqn. \ref{eq:wave_velocity} must be a solution of the continuity Eqn. \ref{eq:continuity_lin} we obtain that the velocity must be perpendicular to the wave vector:

\begin{equation}
\VF{U}\cdot \VF{k}=0 \; \Rightarrow k_z=0
\end{equation}

Therefore, the velocity wave defined by Eqn. \ref{eq:wave_velocity} is transverse, and its wave vector is directed horizontally, $\VF{k}=(k ;0)$.

By imposing that Eqns. \ref{eq:wave_density} and \ref{eq:wave_pressure} are solutions of Eqns. \ref{eq:momentum_lin} and \ref{eq:density_lin} we get:

\begin{eqnarray} 
\label{eq:momentum_fourierx}
k\,P=0  \Rightarrow P=0\\
\label{eq:momentum_fourierz}
i\,\omega \, \rho \, U= g \,R\\
\label{eq:density_fourier}
i \, \omega \, R = \frac{\dif \rho}{\dif z} \, U
\end{eqnarray}

Equation \ref{eq:momentum_fourierx} implies that the pressure does not propagate and the dissipation of energy does not occur through the emission of a sound wave. Multiplication of Eqn. \ref{eq:momentum_fourierz} and \ref{eq:density_fourier} allows to determine the frequency of the internal gravity waves, which coincides with the Brunt-V\"{a}is\"{a}l\"{a} frequency of the oscillating parcel of fluid:

\begin{equation}
\omega= \sqrt{-\frac{g}{\rho_0}\frac{\partial \rho}{ \partial z}}=N  
\label{eq:brunt_freq}
\end{equation}

Notice that, although this theoretical model confirms that the harmonic motion of a parcel of fluid in the vertical direction gives rise to transverse internal gravity waves propagating horizontally, the wave number and the velocity of the wave are not predicted by it, and in principle, there are no constraints on the values that $k$ and the velocity can assume.

\section{Results and discussion}

We performed various experiments using different amounts of salt, ranging between 10 and 46 g/l. 
In each measurement, we tracked the cork's rise until its position was approximately at half of the water height. Then, we took movies of the oscillations in order to measure the Brunt-V\"{a}is\"{a}l\"{a} frequency. 

\begin{figure}[h!]
	\centering
	\includegraphics[width=10 cm]{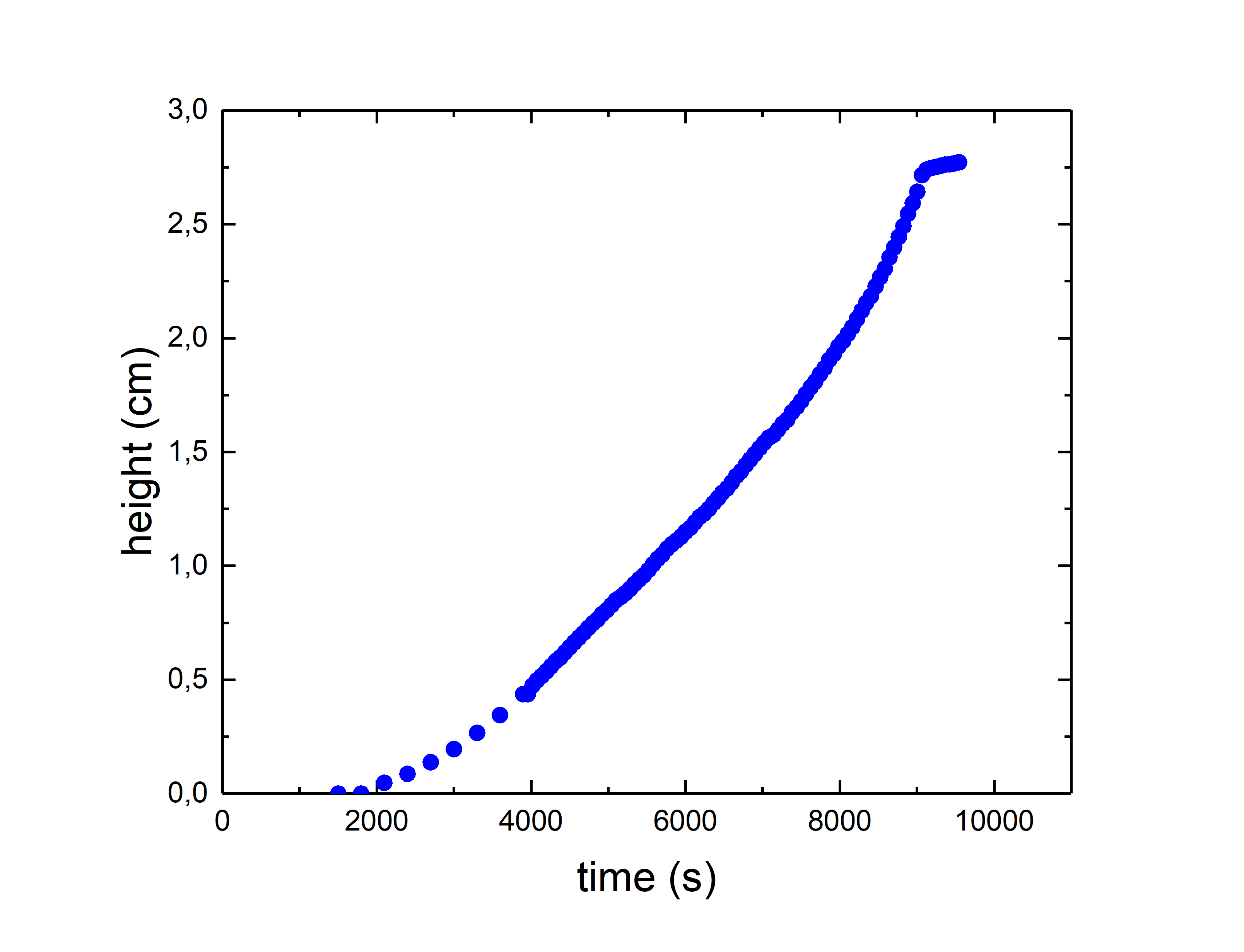}
	\caption{Cork's rise obtained by tracking the position of the marked point  shown in Fig. \ref{fig:cork_rise}.}
	\label{fig:cork_rise2}
\end{figure}

In Figure \ref{fig:cork_rise} we show a sequence of images taken when the cork had a low density. Thanks to the relatively large speed of the process -- it lasted approximately three hours -- it was possible to follow the rise until the cork reached the water surface. Both the marked point and the reference system used with Tracker are shown. In Figure \ref{fig:cork_rise2} the corresponding plot of the position of the cork versus time is shown. In the beginning, shots were taken every five minutes but, due to the high rise speed, the time intervals were soon reduced to one minute. 
The trend shown in Fig. \ref{fig:cork_rise2} exhibits the same features observed in all cases, also when the cork density was increased. There is an initial latency time during which the cork starts to randomly slip at the bottom of the jug, then the rise starts and the height increases almost linearly in time. What is extremely different from one measure to another, and not easily predictable, are the typical latency times, which do not seem to depend trivially on the salt amount poured in water. There are probably multiple reasons for this behaviour. One is that the cork's density can slightly vary from one measure to the other. 

It is also very difficult to guarantee that the salt fills uniformly the bottom of the jug. Moreover, the finite size of the cork is not negligible with respect to the size of the jug. Therefore, the presence of the cork itself perturbs salt dissolution and therefore the typical times of the process. Finally, the temperature of the water is not controlled at all, therefore a change in the transport properties of the mixture is to be expected from one measure to the other.

It is interesting to focus on the energy balance of the system in order to understand the relationship between the different mechanisms acting on the fluid and the cork.
If one writes down the total mechanical energy of the cork as the sum of its gravitational potential energy plus its kinetic energy, when the cork is lifted from the bottom of the jug, the conservation of mechanical energy is obviously not satisfied, since the cork potential energy increases without any reason.
In order to solve this apparent energy mystery, one should seek the origin of the force lifting the cork. The cork is subjected to the weight force, which is constant in time, and to the buoyancy determined by the surrounding liquid, which changes in time. The work needed to raise the cork is provided by the gradual increase of the density of the liquid surrounding the cork, determined by the diffusive dissolution of the salt. Therefore, the cork rises because the center of mass of the fluid surrounding it gradually shifts upwards. This leaves us with another open question: what is the origin of the mechanical energy needed to lift the center of mass of the fluid? The dissolution of salt provides an energy contribution to the system in terms of enthalpy and is driven by the thermal agitation of the component molecules $k_BT$. The energy is, in the end, taken from the thermal energy of the system, so that an accurate measurement of the fluid temperature should reveal a decrease of the temperature during the rise of the salt and the cork. This would, of course, be possible only in an ideal adiabatic system not exchanging energy with other systems. In the experiment described here, the jug and the water are actually in thermal equilibrium with the surrounding environment, and the temperature increase is then not measurable since the environment acts as the ultimate thermal reservoir.
Nevertheless, it is interesting to observe that in the initial condition, with the salt and the cork at the bottom of the jug, the potential energy of the system is at a minimum, while after some time, both the salt and the cork have been lifted increasing their potential energy at the expense of the enthalpy of the salt solution and ultimately to thermal energy. This is one of the  cases of conversion of thermal energy to mechanical energy.

\begin{figure}[h!]
	\centering
	\includegraphics[width=20 cm]{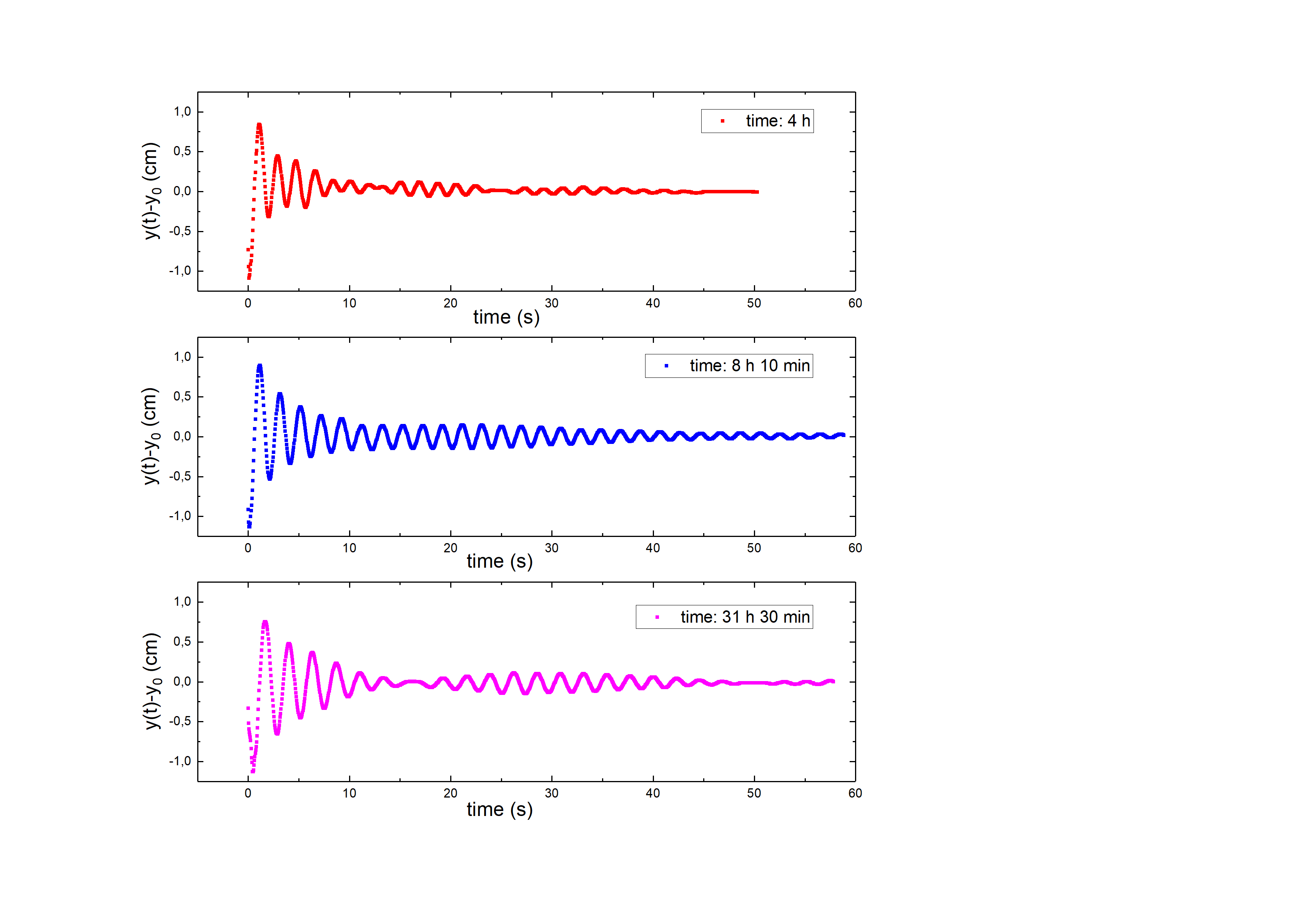}
	\caption{Tracking of the cork oscillation taken at different times of the same measure with 30 g/l NaCl.}
	\label{oscillation2}
\end{figure}

\begin{figure}[h!]
	\centering
	\includegraphics[width=8cm]{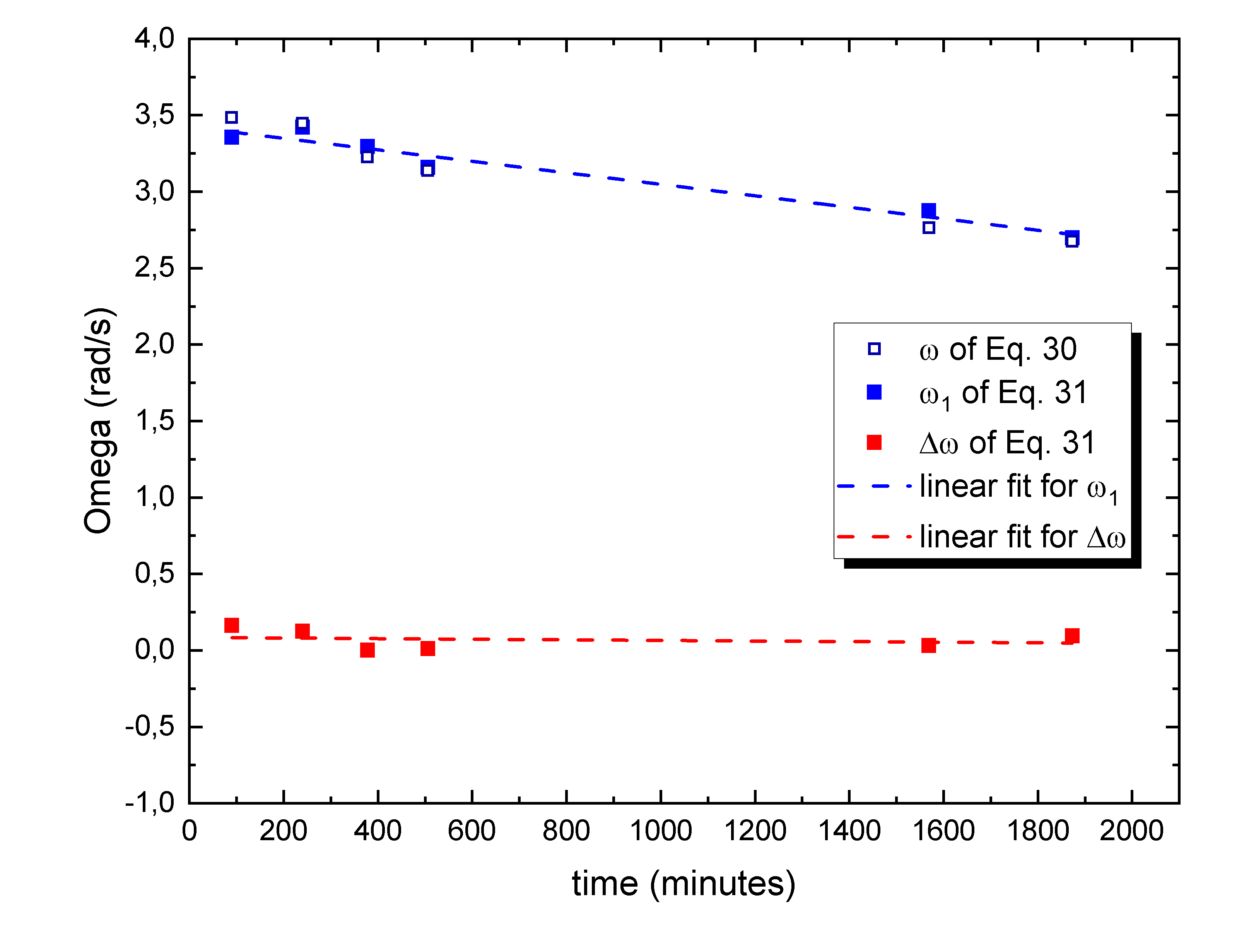}
	\caption{ The plot shows the values of $\omega_1$ (blue squares) and $\Delta\omega$ (red squares), as obtained by fitting the data in Fig. \ref{oscillation2} with Eqn. \ref{eq:fitoscill2}. Open squares are the values of  $\omega$ obtained by fitting the first oscillations of the same data with  Eqn. \ref{eq:fitoscilldamped}. Both $\omega$ and $\omega_1$ give a good estimation of the Brunt-V\"{a}is\"{a}l\"{a} frequency.}
	\label{fig:freq_time}
\end{figure}

At any time during its rise, the cork floats in a stratified medium and, if set in motion with a slight tap, it starts to oscillate at the Brunt-V\"{a}is\"{a}l\"{a} frequency that is directly related to the local density gradient and gravity as shown in Eqn. \ref{eq:brunt}.
In Fig. \ref{oscillation}, we show a sequence of shots of the cork that oscillates after a slight tap, given when it had reached approximately half height within the liquid.
At a visual inspection, it was evident that the regular oscillations of the cork were periodically slowed down until the cork almost stopped, to start again to oscillate after a while. A video is included as supplemental material (video1).
In Fig. \ref{oscillation2} we show a sequence of tracking of the oscillations taken at different times during the cork rise for a slow process where we used the cork with its final density.
The periodical regular slowing down of the cork that can be observed in almost all the figures follows a complex dynamics, which cannot be trivially related to that of a damped harmonic oscillator. However, by fitting the first oscillations with the formula for a damped harmonic oscillator

\begin{equation}
y(t)= A*cos(\omega*t)*exp(-(t/ \tau))+y_0
\label{eq:fitoscilldamped}
\end{equation}

one gets a typical oscillation frequency $\omega$ in the range from 2.67 to 3.45 $rad/s$ and a damping constant $\tau$ from 2.38 to 4.87. The values of the frequency resulting from this simple fit are reported in Fig.~\ref{fig:freq_time}.

A more accurate scrutiny of Fig. \ref{oscillation2} reveals that the dynamics of the oscillations is similar to that observed for a  beating phenomenon that occurs when two waves of slightly different frequencies interfere. 
As discussed at the end of Section 4.3, the oscillations due to the gradient of density are expected to damp emitting transverse internal gravity waves that propagate perpendicularly to the direction of oscillation. Due to the finite size of the container, we expect that these waves are reflected back by the jug walls and finally interfere with the cork oscillation.

\begin{figure}[h!]
	\centering
	\includegraphics[width=16cm]{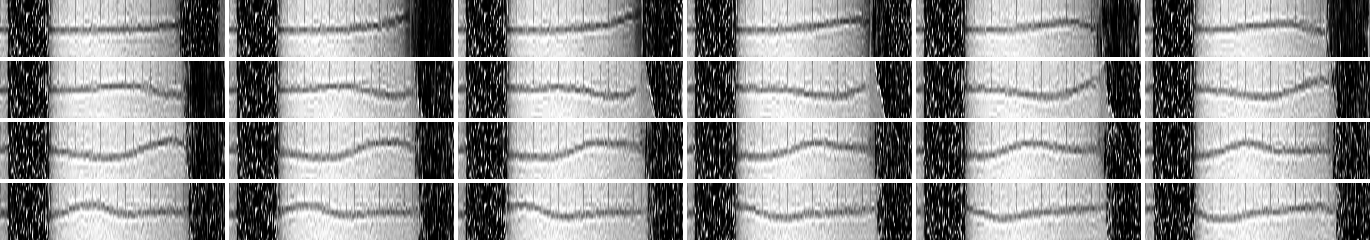}
	\caption{Sequence of images taken with a frame rate of 8 img/s at different times of a portion of the sheet of paper placed behind the tank between the two corks. Time increases from left to right and from top to bottom. The horizontal line is deformed due to the variation of the refractive index produced by the transit of the internal gravity wave. A video is available as supplemental material (video3)}
	\label{squares}
\end{figure}

\begin{figure}[h!]
	\centering
	\includegraphics[width=10 cm]{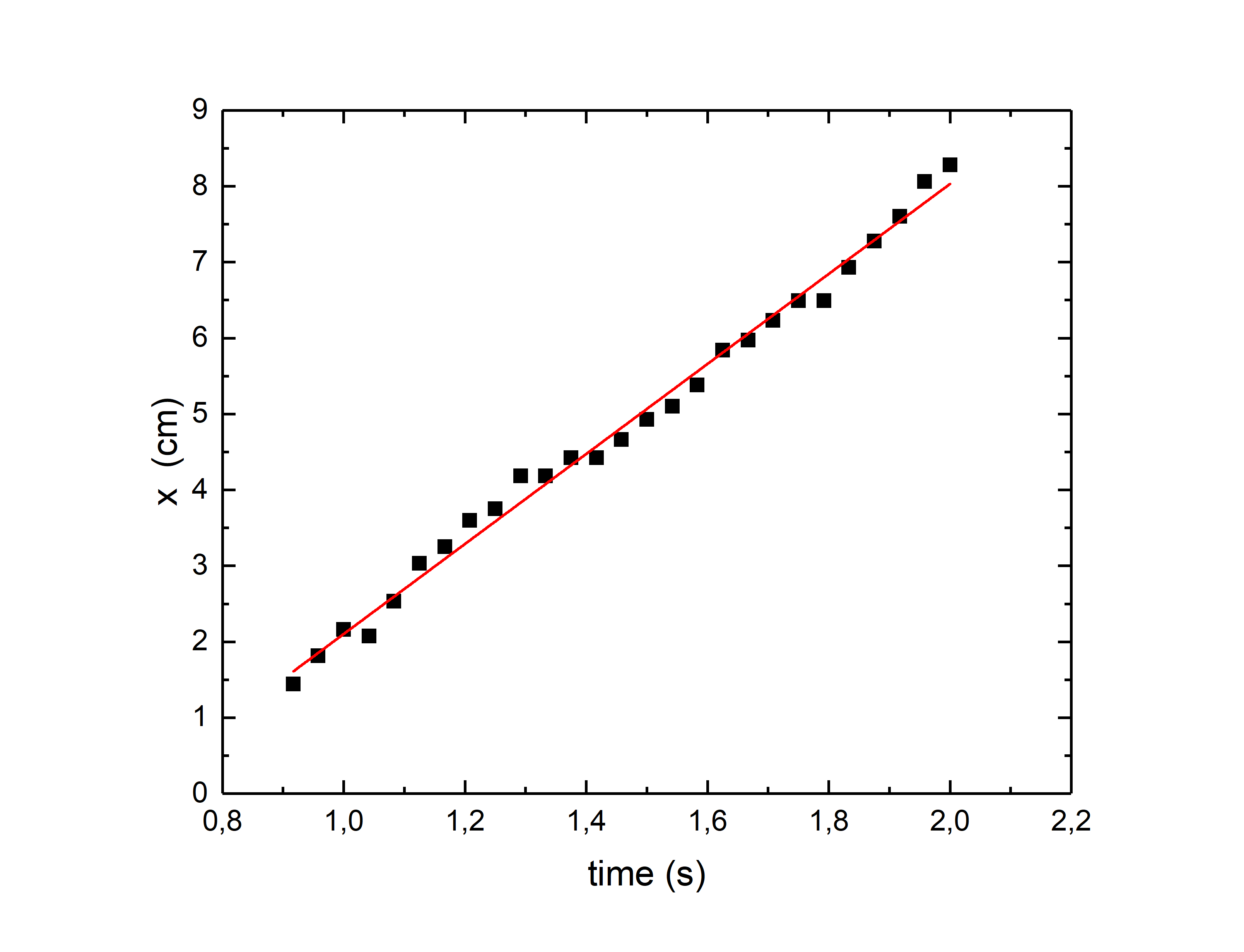}
	\caption{Peak position of the wave detectable as a deformation of the line shown in Fig. \ref{squares}. The solid line is the best linear fit of the experimental points and its slope gives an estimate of the wave speed that, for this curve is  $6,0 \pm 0,1 \ cm/s$.}
	\label{onde}
\end{figure}

To test this hypothesis we performed a further experiment using a rectangular plexiglass tank as a container. We prepared another cork similar to the first one and immersed both of them in water and salt. 
As soon as the corks had reached a proper height, we slightly pushed one of them and verified that after a while also the other one started to oscillate. This is a piece of evidence that a signal propagated horizontally from the first cork to the second, as expected for internal gravity waves. In Figure \ref{Two_corks} an image of the two corks is shown, while a video is included as supplemental material (video2).
Gravity waves in a transparent fluid are almost invisible, and this feature makes them hard to investigate, but some of their features can be determined from their effect on bodies immersed in the fluid, such as the two corks that we used. A remarkable example is represented by the velocity of the wave, which can be obtained from the ratio $\frac{d_c}{t_m}= v = 6,6 \pm 1,4 \ cm/s$, where $d_c = 8,3 \pm 0,5 \ cm$ is the distance between the two corks, measured from the images, and $t_m \simeq 1,25 \ \pm 0,2 s$ is the time interval between the push on the first cork and the time at which the second one starts to oscillate itself. A more thorough investigation of the process requires the possibility of visualizing the wave. This is made possible by the fact that the waves occur through density variations, which are associated to changes of the index of refraction that could be visualized by observing the deformations of a reference object placed behind the liquid. In this experiment, we used a chequered sheet placed in contact with the back window of the tank. 
To analyse the gravity waves, we processed the images of the two corks with Fiji by expanding the images in the vertical direction by a factor of 16 to achieve better visualization of the horizontal lines of the chequered reference pattern. Images were then converted to grayscale and a gamma correction of exponent 3 was applied to increase the contrast of the horizontal lines. A video is available as supplemental material (video3).
In Fig. \ref{squares} we show a sequence of processed images collected at different times after the start of oscillations of the first cork. The deformation of the horizontal line can be clearly observed and used to calculate the speed and frequency of the internal gravity wave. 
The positions of the wave peaks were followed in time with the Tracker program, and in Fig. \ref{onde}, a plot of the displacement along the x-direction vs. time for one of them is shown. The continuous straight line is the best linear fit of the experimental points and its slope gives an estimate of the wave speed $v_w \approx 6,0 \pm 0,1 cm/s$.

We have tracked three different peaks spanning the space between the two corks obtaining an average value for $v_w$, $v_{av}= 6,7 \pm 0,6 \ cm/s$.

\begin{figure}[h!]
	\centering
	\includegraphics[width=8cm]{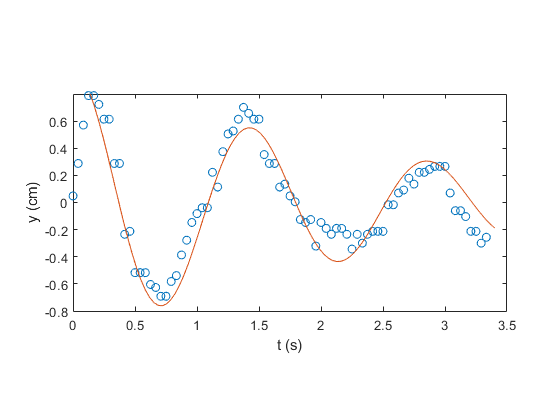}
	\caption{Tracking of the position of a point on the line shown in Fig. \ref{squares} during the passage of the gravitational wave.  The continous line is the fit obtained using Eqn. \ref{eq:fitoscilldamped}.}
	\label{fig:squaresfit}
\end{figure}

We also measured the frequency of the internal gravity wave by using the line deformations along the vertical direction, induced by the index of refraction inhomogeneities shown in Fig. \ref{squares}. Using the Tracker program, fixed points of the line were chosen and followed in time during the passage of the wave to measure their displacement along the y-axis. A typical result is shown in Fig. \ref{fig:squaresfit}.

The continuous line is the best fit of the experimental points with Eq. \ref{eq:fitoscilldamped}.
By averaging the oscillation frequencies obtained by tracking different points of the line, we obtained an average value for $\omega = 4.4 \pm0.4 \ rad/s$ that corresponds to a frequency $f= 0,64 \pm 0.06 Hz $ for the internal gravity wave.
This value is close to the oscillation frequency of the cork pushed in motion, $\omega_c= 4.26 \ rad/s $, as measured by tracking the cork position. This result confirms that the internal gravity wave has the same frequency as the Brunt-V\"{a}is\"{a}l\"{a} wave that generates it, as shown in Eqn. \ref{eq:brunt_freq}, although it must be stressed that the image quality and the strong damping of the waves make it difficult to obtain a precise estimation of the frequencies.

\begin{figure}[h!]
	\centering
	\includegraphics[width=8cm]{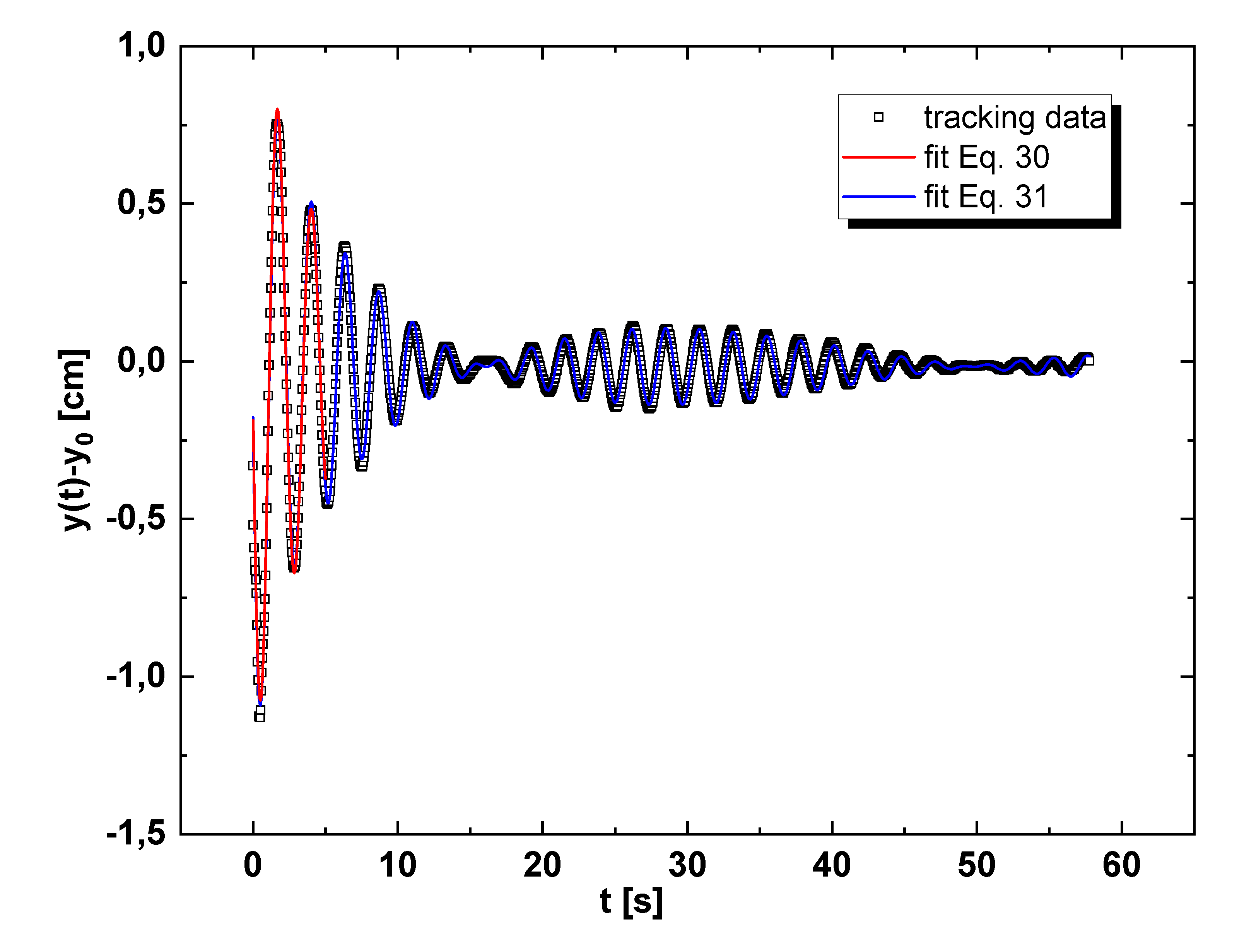}
	\caption{Fit of the oscillation tracking for a measurement performed with 30 g/l NaCl. Data are taken 4 hours after the beginning of data acquisition, when the cork was almost 1,5 cm from the bottom of the jug.}
	\label{fig:fitoscill}
\end{figure}

To have better quality measurements of Brunt-V\"{a}is\"{a}l\"{a} frequencies in different conditions, we finally fitted the individual cork's oscillations shown in Figs. \ref{oscillation} and \ref{oscillation2}. In order to fit the entire curves and not only the first oscillations, we used a phenomenological fitting function more refined with respect to Eq. \ref{eq:fitoscilldamped} that takes into account both the oscillation damping and the beatings between the cork oscillation and the back-reflected internal gravity wave:

\begin{equation}
y(t)= A*cos(\omega_1*t+\phi _1)*cos(\Delta\omega * t+ \phi _2)*exp(-(t/ \tau)^\gamma)+ y_0
\label{eq:fitoscill2}
\end{equation}

where $\omega_1$, $\Delta \omega$, $\phi_1$, $\phi_2$, $\tau$, $\gamma$ and $y_0$ are the free parameters. Eq. \ref{eq:fitoscill2} is the product of two cosine functions which typically describes beatings that occur when two waves of similar frequencies interfere. In beating phenomena, $\omega_1$ is the average of the frequencies of the interfering waves, while  $\Delta\omega$ is their difference. The stretched exponential term $exp(-(t/ \tau)^\gamma)$ with $\gamma =2$ turned out to be the best function to account for the severe damping of the two oscillating terms that have different time constants. In Fig. \ref{fig:fitoscill} we show the remarkable result of the fit. It can be noticed that while Eqn. \ref{eq:fitoscilldamped} is able to fit only the first oscillations, Eqn \ref{eq:fitoscill2} is able to account for all the oscillations features.

As shown in Fig. \ref{fig:freq_time}, for all the data shown in Fig. \ref{oscillation2}, $\Delta\omega$ is much smaller than $\omega_1$ and this result confirms that the two beating waves have very close frequencies. As a consequence, $\omega_1$ gives a good estimation of the Brunt-V\"{a}is\"{a}l\"{a} frequency and turns out to be in very good agreement with the frequency $\omega$ that is obtained by fitting the data with the much simpler Eqn. \ref{eq:fitoscilldamped}. In Fig. \ref{fig:freq_time}, it can be appreciated that the frequency of oscillation decreases in time as expected because the density gradient diminishes while the concentration approaches the final equilibrium value.

From the Brunt-V\"{a}is\"{a}l\"{a} frequency, it is possible to calculate the value of the density gradient  $\frac{\partial \rho}{ \partial z}$ from Eqn. \ref{eq:brunt}.
Having fitted all the oscillations measured with all the different salt concentrations, we found for the frequency $N= \omega$ values between $1,5$ and $4,5$ rad/s depending on the initial salt concentration and the hight of the cork in the fluid and on the time of the measurement. Using Eqn. \ref{eq:brunt} the corresponding values for $\frac{\partial \rho}{ \partial z}$ ranges from $236 \ kg/m^4$ to $2140 \ kg/m^4$ corresponding to concentration variations $\frac{\partial c}{ \partial z}= \frac{\partial \rho}{ \partial z}/\frac{\partial \rho}{ \partial c}$ ranging from $0.32 \ m^{-1}$ to $3.0\ m^{-1}$, where for a solution of NaCl in water at small concentration $\frac{\partial \rho}{\partial c}=720 kg/m^4$ \cite{crc2005}.

\section{Conclusions}

We have discussed an engaging, inexpensive, and simple experiment that allows introducing undergraduate students to many arguments on hydrostatic, buoyancy, and levitation in a stratified fluid. 
The general scientific framework of levitation, and the theoretical description of the phenomena that are experimentally accessible, are also presented in this work.
Brunt-V\"{a}is\"{a}l\"{a} oscillations are observed and the measurement of their frequency may be used to estimate the concentration gradient inside the fluid. Moreover, the effect of gravity waves can be observed and measured. We propose a phenomenological function to fit the data, able to keep into account both the damping of the oscillations and the beatings with the internal gravity wave they generate.
The motion of the cork determines a partial remixing of the stratified fluid, thus reducing the concentration gradient that drives the oscillations. In this respect, the experiment represents a remarkable example of a classical system where performing a measurement significantly affects the state of the system. This feature makes a long series of repeated measurements not useful, in contrast with the usual physical systems usually encountered by students during laboratory activities.
The experiment is exploitable by students of any age as the level of deepening can be tuned as a function of the students' knowledge. However, it is particularly suitable for undergraduate students that can appreciate all its implications.

\section*{Acknowledgements}
Work partially supported by the European Space Agency, CORA-MAP TechNES Contract No. 4000128933/19/NL/PG.
This research was carried under the framework of the E2S UPPA Hub Newpores and Industrial Chair CO2ES supported by the ‘Investissements d'Avenir’ French programme managed by ANR (ANR–16–IDEX–0002).

\section*{References}

\bibliography{corknew6}

\end{document}